# Coherent coupling of a single spin to microwave cavity photons

J. J. Viennot*, M. C. Dartiailh, A. Cottet and T. Kontos*

Laboratoire Pierre Aigrain, Ecole Normale Supérieure-PSL Research University, CNRS, Université Pierre et Marie Curie-Sorbonne Universités, Université Paris Diderot-Sorbonne Paris Cité, 24 rue Lhomond, 75231 Paris Cedex 05, France

*Correspondence to:  viennot@lpa.ens.fr ; kontos@lpa.ens.fr

Electron spins and photons are complementary quantum-mechanical objects that can be used to carry, manipulate and transform quantum information. To combine these resources, it is desirable to achieve the coherent coupling of a single spin to photons stored in a superconducting resonator. Using a circuit design based on a nanoscale spin-valve, we coherently hybridize the individual spin and charge states of a double quantum dot while preserving spin coherence. This scheme allows us to achieve spin-photon coupling up to the MHz range at the single spin level. The cooperativity is found to reach 2.3, and the spin coherence time is about 60ns. We thereby demonstrate a mesoscopic device suitable for non-destructive spin read-out and distant spin coupling.

The methods of cavity quantum electrodynamics hold promise for an efficient use of the spin degree of freedom in the context of quantum computation and simulation *(1)*. Realizing a coherent coupling between a single spin and cavity photons could enable quantum non-demolition readout of a single spin, quantum spin manipulation, and facilitate the coupling of distant spins *(1, 2, 3, 4)*. It could also be used in hybrid architectures in which single spins are coupled to superconducting quantum bits *(5),* or to simulate one-dimensional spin chains *(6)*.

The natural coupling of a spin to the magnetic part of the electromagnetic field is weak *(7)*. In order to enhance it, one needs a large spin ensemble, typically of about $10^{12}$ spins *(8, 9, 10, 11, 12, 13)*, but these ensembles lose the intrinsic non-linearity of a single spin 1/2. Alternatively, several theoretical proposals have been put forward to electrically couple single spins to superconducting resonators in a mesoscopic circuit *(14, 15, 16, 17)*, building on the exquisite accuracy with which superconducting circuits can be used to couple superconducting qubits and photons and manipulate them *(18)*. One such approach is to engineer an artificial spin-photon interaction by using ferromagnetic reservoirs *(15)*. Noteworthy, the spin/photon coupling is also raising experimental efforts in the optical domain *(19,20,21,22,23)*, but the circuit approach presents the significant advantage of scalability.

Recent experiments have demonstrated the coupling of double quantum dot charge states to coplanar waveguide resonators, with a coupling strength $g_{charge} \approx 2\pi \times 10 - 50$ MHz *(24, 25, 26, 27, 28)*. In Ref, *(29)*, the spin blockade read-out technique in quantum dots *(30)* was combined with charge sensing with a microwave resonator *(31)*. In contrast to this spin-blockade scheme, here we use the ferromagnetic proximity effect in a coherent conductor to engineer a spin-photon coupling. Our scheme relies on the use of a non collinear spin valve geometry, which realizes an artificial spin orbit interaction *(15)*. Specifically, we contact two non collinear ferromagnets on a carbon nanotube double quantum dot.

Our device is shown in Fig. 1, A-C. Our resonator is similar to a previous experiment *(27)* with a coupling scheme adapted from *(24)*. It is a Nb resonator with a quality factor $Q \approx$ 6200 – 11800, depending on external magnetic field (see fig. S6). We use a previously developed technique of stamping in order to preserve the Q factor of the resonator, and use nanotubes grown by chemical vapor deposition *(32)*. The imprints of the stamps used to transfer the nanotubes are visible in Fig. 1A. We use PdNi as a ferromagnetic alloy. It forms good contacts with CNT's *(33, 34)* and its magnetization direction is simply controlled by the geometry of the electrodes *(35)*. We set an angle (45° at zero magnetic field) between the magnetizations of the electrodes.

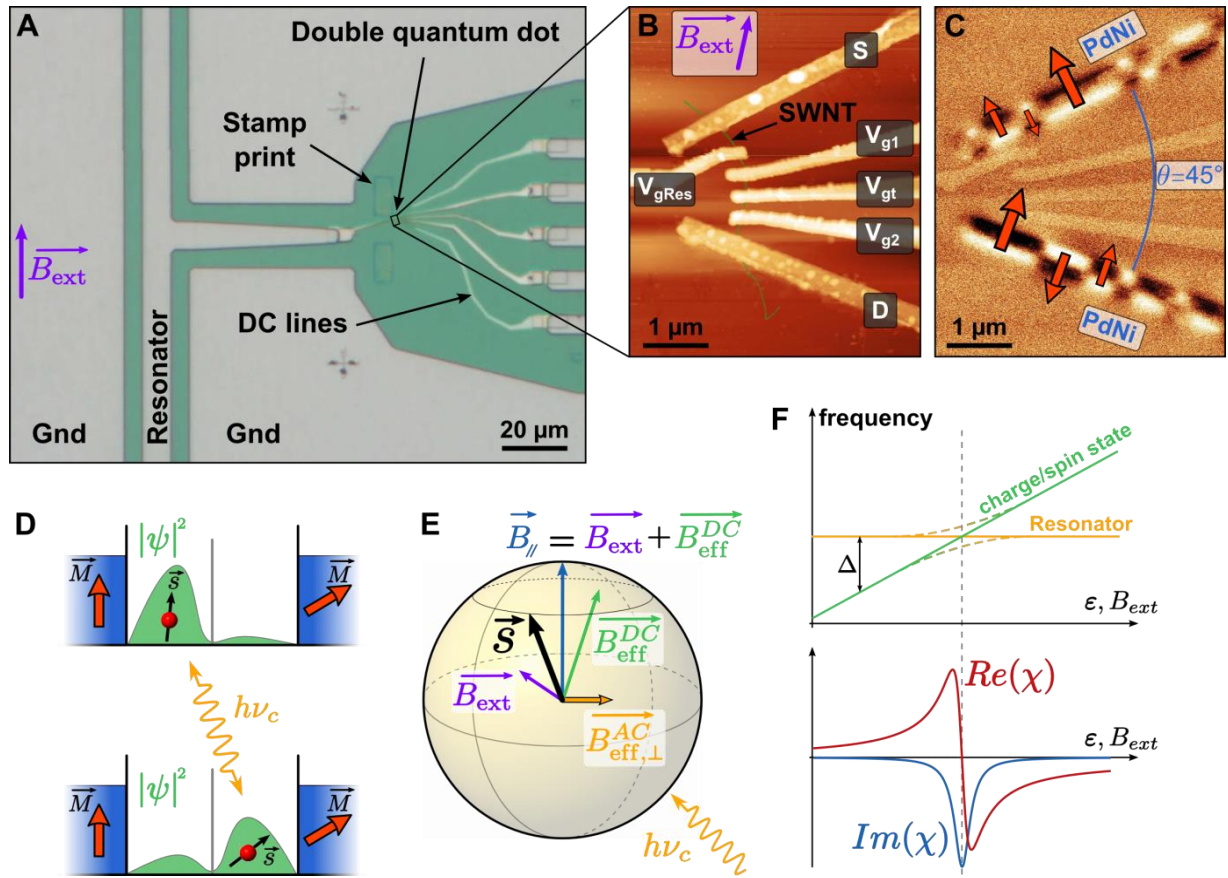

**Figure 1. Experimental setup.** **(A)** Optical micrograph of the essential part of the device. The resonator central conductor is surrounded by ground planes which are open in a small region, in order to place a carbon nanotube and the necessary DC lines to form a double quantum dot. **(B)** Atomic force micrograph of the nanotube with four top gates used to bring DC electrical potentials as well as couple to the resonator via $V_{gRes}$. As shown in the magnetic force micrograph **(C)**, source S and drain D electrodes are made out of a ferromagnetic alloy (PdNi). Black and white colors correspond to north and south poles of ferromagnetic domains.

**(D)** General principle of our coupling mechanism. The proximity of the non collinear ferromagnets induces a different equilibrium spin orientation if an electron is localized in the left or in the right dot. Photons are coupled to transitions changing the localization of the wavefunction Ψ, and hence coupled to transitions changing the spin orientation. **(E)** Bloch sphere of the electron spin showing the different magnetic fields contributions. **(F)** When a transition crosses – or anti-crosses – the resonator frequency, the associated susceptibility exhibits a resonance that is directly observable in the phase and amplitude of the resonator transmission

Ferromagnets deposited on carbon nanotubes induce effective magnetic fields *(33, 36)*. In our setup, each dot is contacted to one ferromagnet creating a local effective magnetic field and therefore a Zeeman splitting with a quantization axis given by the ferromagnet magnetization direction. When an electron is moved from one dot to the other, its equilibrium spin orientation rotates (Fig. 1D). As a consequence, we obtain an artificial spin-orbit coupling, engineered extrinsically. The localization of an electron wavefunction depends on the inter-dot energy detuning ε *(31)*. This parameter is controlled experimentally with DC voltages applied with local top gates. Importantly, because this control parameter is governed by electric fields, it is also actuated by the AC electric field associated with the photons in the resonator *(24, 25, 26, 27, 28, 29)*. A single electron spin is thus coupled to the photons of the resonator via the natural coupling of the double dots' charge orbitals to the resonator electric field (Fig. 1D). In order to tune the transition frequency of the spin states and bring them in resonance with the cavity frequency, we also apply an external magnetic field $B_{ext}$. The different contributions of real and effective magnetic fields are depicted in Fig. 1E. The longitudinal component $B_\parallel$ is given by the sum of the external magnetic field and the DC part of the effective fields, which depends on ε. This controls the Larmor frequency of the spin.

When the electric field of the cavity actuates a charge modulation between the two dots, an AC effective magnetic field appears, with a transverse component $B^{AC}_{eff,\perp}$ oscillating at the resonator frequency. This yields the two non collinear magnetic fields necessary for transverse coupling to the spin: a DC longitudinal component, and a transverse AC component that can be oscillating at the Larmor frequency. Note that although the hybridization of the spin with the charge orbitals is the mechanism responsible for coupling to the electric field of the resonator, in principle this hybridization can be weak enough to preserve spin coherence *(15)*. Moreover, the sensitivity to charge noise depends on the dispersion relation of the hybridized spin-charge transitions, and therefore it can have a different behavior than the spin-photon coupling strength. By sweeping ε, the resonant lines in Fig. 2, A and B go through a sweet spot where the double quantum dot (DQD) transition frequencies are minimum and therefore insensitive to charge noise at first order. The existence of this 'sweet spot' contributes to the high cooperativity found for the spin transitions.

We measure the amplitude A and phase φ of the cavity transmission at resonance, at 40mK. We tune the gate voltages of the device to form a DQD (see fig. S1). Transitions in the DQD yield phase and amplitude shifts Δφ and ΔA/A of the resonator transmission. The intrinsic dependence of the superconducting cavity on external magnetic field is taken into account in all measurements. For every change in $B_{ext}$, the reference phase and amplitude are measured first, in order to obtain the correct Δφ and ΔA/A *(37)*. A given transition between two DQD states |i⟩ and |j⟩ is characterized by a susceptibility to a microwave excitation $\chi_{ij}$ $= \dfrac{g_{ij}}{-i\Gamma_{ij}/2 + \Delta_{ij}}$, where $g_{ij}$ is the coupling strength, $\Gamma_{ij}$ is the decoherence rate and $\Delta_{ij}$ is the frequency detuning *(37)*. In practice, when such a transition is brought into resonance with the cavity mode, it shifts the mode frequency $f_c$ by $\text{Re}(\chi_{ij})$ and changes the mode linewidth $\kappa$ by $\text{Im}(\chi_{ij})$. The general form $\chi_{ij}$ is plotted in Fig. 1F. This signal is encoded in $\Delta\varphi$ and $\Delta A/A$ which, to first order in $\chi_{ij}$, are respectively given by $\text{Re}(\chi_{ij})$ and $\text{Im}(\chi_{ij})$. In the presence of multiple transitions, the phase and amplitude shifts are given by the sum of all the susceptibilities associated with transitions starting from a populated energy level *(8, 24)*.

Figures 2,A and B show, respectively, the phase and amplitude shifts as functions of inter-dot gate detuning $\varepsilon$ and external magnetic field $B_{ext}$. Sign changes are observed in $\Delta\varphi$, together with dips in $\Delta A/A$, indicating DQD transition frequencies crossing the cavity frequency. Figure 2C shows a line cut at $B_{ext} = 59\text{mT}$. The variation of the phase and amplitude in the dashed area of Fig. 2C resembles Fig. 1F. Figures 2,A and B therefore demonstrate that three transitions of the DQD are coupled to the cavity, and disperse as functions of both $\varepsilon$ and $B_{ext}$. This dependence on both gate voltage and magnetic field strongly suggest transitions involving changes in both charge and spin states (Fig. 2D). Charge states in the dots, separated by the energy $\varepsilon$, are Zeeman split by the effective fields induced by the ferromagnets. The tunnel coupling between the two dots coherently hybridizes their orbitals to form the analog of bonding and anti-bonding states *(27, 31)*. Besides, the non collinear quantization axis of the two dots couples the spin populations. The four states of Fig. 2D thus coherently hybridize into four quantum states having both charge and spin components.

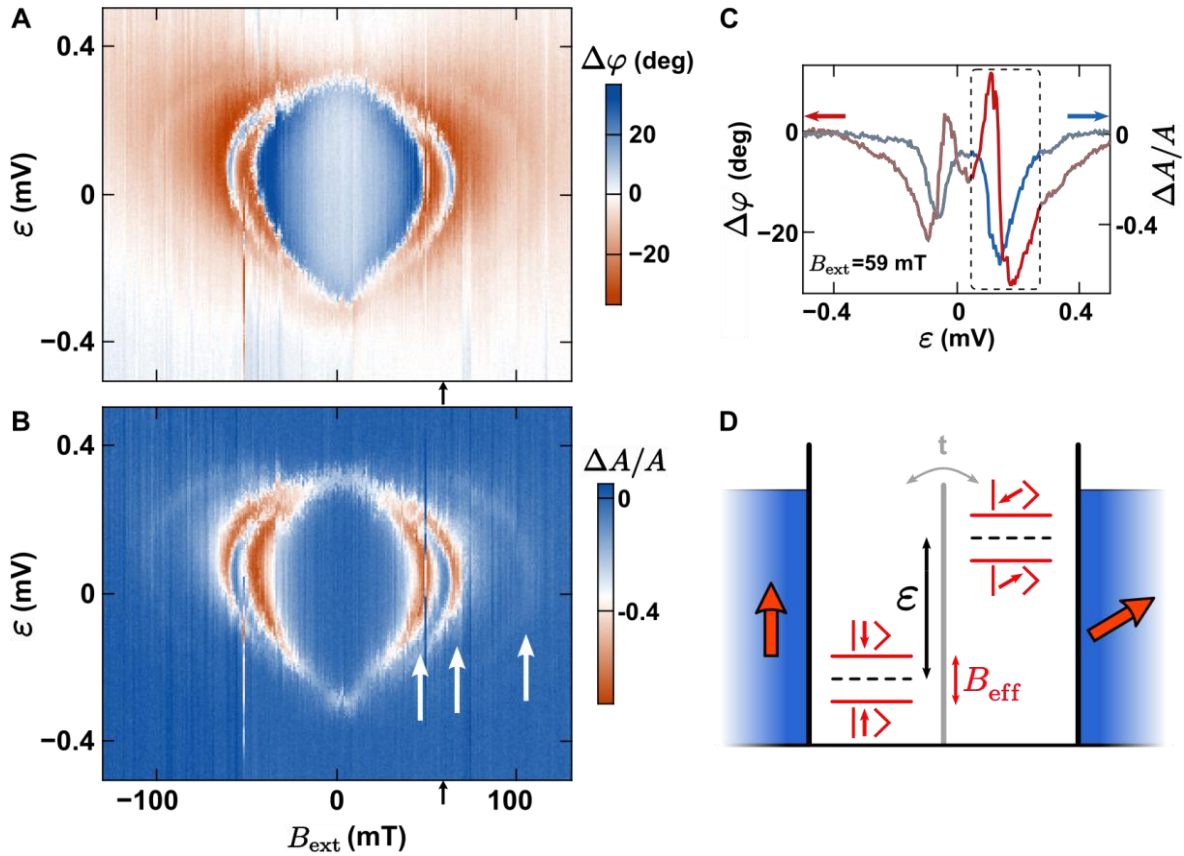

**Figure 2. Electric and magnetic dependence of the quantum dot transitions.** (A) and (B) Measured phase and amplitude signals as a function of external magnetic field and inter-dot gate detuning ε, at a microwave power P ≈ -116dBm, (about 40 photons in the cavity). We identify three transitions indicated by white arrows. The temperature is 40mK. (C) Phase and amplitude versus ε at $B_{ext}$ = 59mT (indicated by black arrows in (A) and (B)) showing resonances similar to Fig. 1F within the dashed region. (D) Charge states (black dashed levels) are spin-split thanks to the effective fields $B_{eff}$. The four states (red levels) coherently hybridize via the inter-dot tunnel coupling t.

When a transition is resonant with the cavity (i.e. around $\Delta_{ij}$=0), the phase and amplitude contrasts are directly linked to the cooperativity:

$$C_{ij} = \frac{4g_{ij}^2}{\kappa \Gamma_{ij}}$$

The cavity linewidth κ is varying slowly and by less than a factor of two in our magnetic field range ($2\pi \times 0.61$ MHz $< \kappa < 2\pi \times 1$ MHz, see fig. S6). Nevertheless we observe a higher phase and amplitude contrast at higher fields and smaller inter-dot detuning ε (see Figs. 2, A and B). This indicates that transitions become more coherent, or more coherently coupled to the cavity in this region.

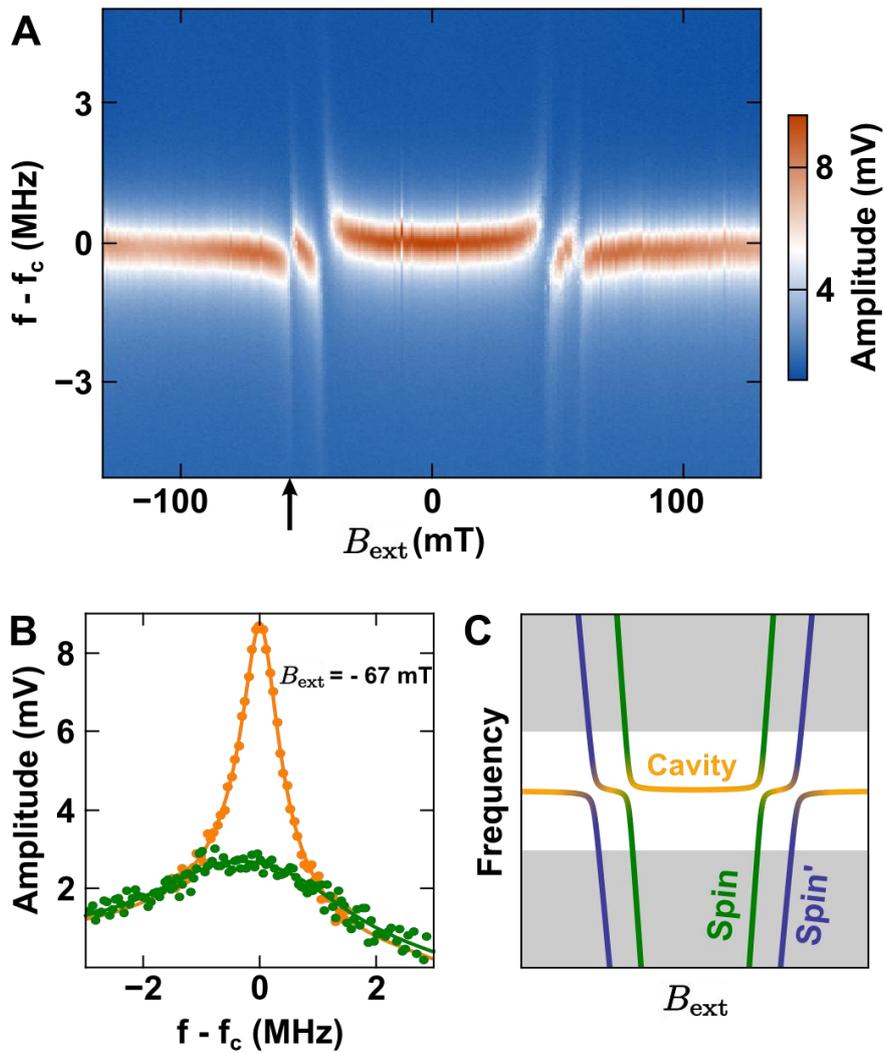

**Figure 3. Spin-cavity hybridization.** **(A)** Measured transmission spectrum of the cavity as a function of $B_{ext}$, centered around the resonant frequency $f_c$, at small detuning $\varepsilon \approx 50\mu V$. The temperature is 40mK. **(B)** Resonator transmission at $B_{ext} = -67mT$ (indicated by the black arrow) with spin transition detuned ($\varepsilon > 1mV$, orange curve) and resonant ($\varepsilon \approx 50\mu V$, green curve), measured at microwave power $P \approx -119dBm$, (about 20 photons in the cavity). Circles are data and solid line is theory explained in *(37)*. **(C)** Sketch of the spin transitions (Spin and Spin') dispersing with $B_{ext}$ and hybridizing with the resonator mode (Cavity). Our measurements focus on the white stripe around the cavity frequency.

In order to be more quantitative, we measure the resonator transmission spectrum as a function of magnetic field at small inter-dot detuning $\varepsilon$ of about 50μV (Fig. 3A). For the sake of clarity, the frequency traces are all centered around the bare cavity frequency $f_c$, which itself shows jumps with changes in $B_{ext}$ (see fig. S2 for the data in absolute frequency). At ±50mT two DQD transitions become resonant with the cavity and cause strong distortions on the transmission spectrum. This confirms the high cooperativity between spin hybridized transitions and the resonator. Figure 3B shows profiles of the resonator transmission at -67mT, for a strongly detuned ($\varepsilon > 1mV$, orange curve), and a resonant transition ($\varepsilon \approx 50\mu V$, green curve). We observe a dramatic change in the amplitude and width of the transmission. Fitting this data *(37)*, we extract the bare cavity parameters and estimate the coupling strength $g_{spin} \approx 2\pi \times 1.3$ MHz for this transition, with a decoherence rate $\Gamma^*_{2, spin}/2 \approx 2\pi \times 2.5$ MHz, corresponding to a cooperativity $C \approx 2.3$ *(37)*. This is to be compared to the much larger charge decoherence rate $\Gamma^*_{2, charge}/2 \approx 2\pi \times 0.45 - 3$ GHz measured previously in similar conditions on a carbon nanotube *(27)* and arising from charge noise. For the neighboring transition at -43mT, we find $C \approx 3.3$ which suggests that this transition is also dominantly a spin transition. Figure 3C shows a sketch of the spectrum obtained from a Hamiltonian generalized from *(15)* (see *(37)*). In this sketch, we omit the third (faint) resonance visible in Figs. 2,A and B because of its weaker coherence. The calculated spectrum in *(37)* is in agreement with Fig.3A. We are also able to reproduce the three resonances in Figs.2, A and B (fig. S5). In our model, the two strongest resonances correspond dominantly to spin transitions, as expected. The third faint resonance corresponds to a transition which is less coherently coupled *(37)*.

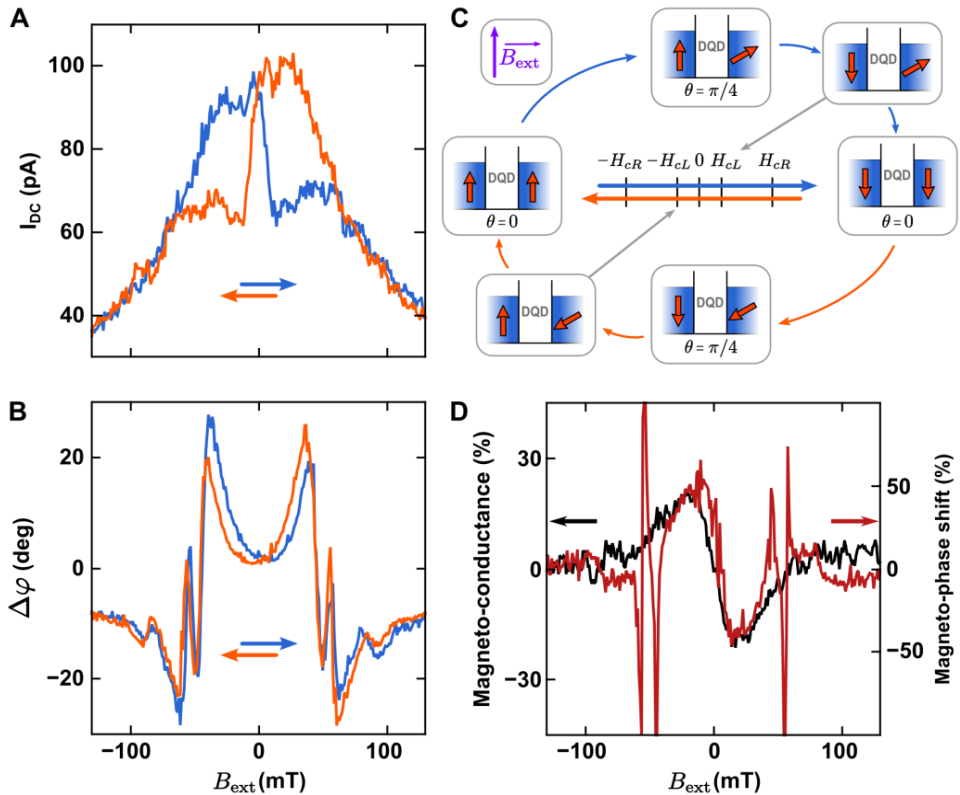

**Figure 4. Magnetic hysteresis.** **(A) and (B)** Measured DC current and microwave phase shift for increasing (blue) and decreasing (orange) external magnetic field. See *(37)* for the details on the phase measurement. The temperature is 40mK. **(C)** Schematics of the ferromagnets

magnetization evolution with $B_{ext}$. $H_{cL}$ and $H_{cR}$ are the coercive fields of the ferromagnets. **(D)** Percentage of hysteresis for the DC current (magneto-conductance) and the phase signal *(37)*.

The final piece of evidence that the hybridized spin states arise from the extrinsic artificial spin-orbit interaction is provided by operating our device as a spin valve. To achieve that, we swept the magnetic field fast enough for the ferromagnetic electrodes to switch hysteretically. Figure 4A shows a conventional DC current measurement as function of the magnetic field and the magnetic field sweeping direction, done on a co-tunneling line (such as shown in fig. S1). We observe the characteristic hysteretic behavior of a quantum dot spin valve, which can be explained by the magnetization reversal sequence of Fig. 4C. This typically results in a magneto-conductance such as the one shown in Fig. 4D. Importantly, the electrodes' magnetizations are non collinear and one of them is not aligned with the magnetic field. The magnetization configuration is therefore non-collinear up to high enough $B_{ext,}$, preserving the amplitude of spin flipping matrix elements in the spin/photon Hamiltonian for transitions Spin and Spin' *(15)*. Figure 4B shows the phase shift close to $\varepsilon = 0$ for the same type of measurement as in Fig. 4A. . The main part of the signal simply corresponds to the profile of Fig. 2A taken at small $\varepsilon$, showing the three transitions. Strikingly, the phase is hysteretic, revealing a hysteresis in the total susceptibility $\chi$. We have ensured that this is not caused by spurious hysteresis of the cavity by measuring systematically all the cavity parameters along the hysteresis path (fig. S6). In Fig. 4D, we plot the hysteretic part of the phase shift *(37)*. Sharp variations correspond to fields where the spin transition is resonant with the cavity. At these points, a small hysteresis in the transition frequency, therefore in the detuning $\Delta_{ij}$, yields a strong hysteresis in the susceptibility. Outside of these areas we observe a smooth variation similar to the behavior of the magneto-conductance. Both magneto-conductance and magneto-phase shift thus vary on the same scale of magnetic field. This is further evidence that the spectrum is genuinely affected by the ferromagnets.

Along with a single spin-photon coupling strength $g_{spin} \approx 2\pi \times 1.3$ MHz, we are able to give a lower bound for the spin decoherence time in carbon nanotubes $T_2^* > 60$ ns ($T_2^*=2/\Gamma_2^*$). This is already almost one order of magnitude larger than the previous measurements in nanotubes *(38)*, but we believe that it could be improved further by optimizing the spin-charge hybridization. The cooperativity and decoherence rates given above indicate that our system is at the strong coupling threshold. Owing to the general principle used here, this method could be applied to many host materials for spin quantum bits. These results open an avenue for single spin-based circuit QED experiments.

**Acknowledgments:** We acknowledge fruitful discussions with the Quantronics group, J. M. Raimond, B. Huard, A. Thiaville, L. Bretheau, E. Flurin, L.E. Bruhat, M.P. Desjardins and technical support of M. Rosticher and J. Palomo. We gratefully acknowledge P. Senellart and L. Lanco for discussions and for communicating results to us prior to publication. The work was financed by the EU project SE2ND and the ERC Starting grant CirQys. The data described in the paper are presented in this report and supplementary materials.

# Supplementary Materials for

## Coherent coupling of a single spin to microwave cavity photons

J.J. Viennot, M.C. Dartiailh, A. Cottet and T. Kontos.

correspondence to: viennot@lpa.ens.fr & kontos@lpa.ens.fr

**This file includes:**

Materials and Methods

Figs. S1 to S6

**Materials and Methods**

Double quantum dot susceptibility at zero magnetic field

Similarly to previous hybrid cavity– quantum dot experiments, two different physical quantities can be measured on the device. We can perform a DC transport measurement, acquiring the current flowing through the double quantum dot as a function of gate voltages. This allows us to obtain a conventional transport spectroscopy as shown in figure S1. We can simultaneously measure the phase and amplitude of the microwave field transmitted through the cavity, which is sensitive to the susceptibility of the double quantum dot transitions. Figure S1 shows such measurements in a gate region where the carbon nanotube device behaves as a double quantum dot. These color-scale plots outline the stability diagram of the device in this region, at zero external magnetic field. We label the charge occupation numbers of the dots by n and m. At the degeneracy between the dot's left/right charge occupation states (n,m+1) and (n+1,m), we observe sign changes in the phase signal along two parallel lines indicating transitions in the double quantum dot that are resonant with the microwave cavity.

Measurement techniques

The microwave measurement techniques are essentially similar to those used in ref *(26)*. We measure the amplitude and the phase of the transmitted microwave signal as a function of the various parameters of the system (frequency of drive, magnetic field or DC gate voltages) either using a heterodyne detection scheme or a Vector Network Analyzer (VNA).

For every change in magnetic field, we first search and measure the bare cavity frequency $f_c$ by strongly detuning all the DQD transitions from the cavity (typically by going to $\varepsilon > 1$mV). We measure the cavity linewidth and the frequency dependence of the phase at maximum transmission (phase slope). The latter gives the phase sensitivity to a resonant frequency change. We acquire precisely the bare (i.e. the reference) phase and amplitude, then we tune the double dot, going back to $\varepsilon$ values given in the data and subsequently measure phase and amplitude shifts from their bare value, namely $\Delta\varphi$ and $\Delta A/A$. This allows us to compensate for the weak dependence of $f_c$ on magnetic field, and for the jumps associated to magnetic flux vortices penetrating the superconducting film (see ref 8).

Fabrication method

A 150nm thick Nb film is first evaporated on an RF Si substrate at rate of 1nm/s and a pressure of $10^{-9}$ mbar. The cavity is made subsequently using photolithography combined with reactive ion etching ($SF_6$ process). Carbon nanotubes are grown with Chemical Vapor Deposition technique (CVD) at about 900°C using a methane process on a separate quartz substrate and stamped onto the device chip in the desired location inside the gap of the cavity *(31)*. The nanotubes are then localized and contacted in two e-beam lithography steps with top gates and PdNi source and drain contacts, which carry DC signals, as shown in figure 1A and 1B. The top gates are a multilayer of 6nm of $Al_2O_3$ covered with 50 nm of Al and 20nm of Pd. The $Al_2O_3$ is obtained in 3 steps by evaporating 2nm of Al and oxidizing this layer by letting 1mbar of $O_2$ for 10 min. The PdNi source drain electrodes are 30nm thick, 150nm wide $Ni_{75}Pd_{25}$ layers capped with a 5nm Pd layer.

Determination of decoherence and spin-photon coupling strength

In order to derive the transmission coefficient of the cavity, we first consider one transition between energy levels i and j in the double dot, coupled to the cavity mode. From the traditional Jaynes-Cummings Hamiltonian, we write the conventional equations of motion :

$$\frac{d}{dt}a = -(\kappa/2 + i\Delta)a + \sqrt{\kappa_1}a_{in,1} + \sqrt{\kappa_2}a_{in,2} - ig_{ij}\sigma_-$$

$$\frac{d}{dt}\sigma_- = -(\Gamma_{ij} + i\Delta_{ij})\sigma_- + ig_{ij}a\sigma_z$$

$$\frac{d}{dt}\sigma_z = -\gamma_{ij}(\sigma_z + 1) - 2ig_{ij}(a\sigma_+ - a^+\sigma_-)$$

where $a$ is the annihilation operator of the cavity field, $a_{in(out),1(2)}$ are annihilation operators for fields propagating inwards (outwards) the cavity at port 1 (2), $\kappa = \kappa_{int} + \kappa_1 + \kappa_2$ is the total cavity decay given by the sum of internal loss and coupling to the two ports of the resonator. Above, $g_{ij}$ is the coupling constant of the transition to the resonator, $\Gamma_{ij}$ its total decoherence rate (relaxation + dephasing), $\gamma_{ij}$ its relaxation rate, and $\Delta_{ij}$ its detuning to the drive frequency. We introduce the Pauli operator $\sigma_z$ associated to the transition, and the lowering and raising operators $\sigma_-, \sigma_+$.

We now write the input-output relation in which we add a small correction arising from the direct parasitic (and weak) transmission channel in our sample holder :

$$a_{out,2} \approx -i\sqrt{T}e^{i\zeta}a_{in,1} + \sqrt{1-T}\,e^{i(\zeta-\theta)}a_{in,2} + \sqrt{\kappa_2}\,a$$

Above, $T$ and $\zeta$ account for the amplitude and phase of the direct parasitic (and weak) transmission channel in our sample holder and $\kappa_2$ is the coupling rate of the cavity to port 2. The $\approx$ sign indicates that such an equation is only valid at lowest order in $T$ and does not ensure unitarity of the scattering matrix for arbitrary values of $T$.

In the semi-classical limit, we make use of the usual decoupling $\langle a\sigma_z \rangle \approx \langle a \rangle \langle \sigma_z \rangle, \langle a\sigma_+ \rangle \approx \langle a \rangle \langle \sigma_+ \rangle, \langle a^+\sigma_- \rangle \approx \langle a \rangle^* \langle \sigma_- \rangle$ and compute the transmission coefficient $S_{21} = \langle a_{out,2} \rangle / \langle a_{in,1} \rangle$:

$$S_{21} = \frac{\alpha}{2\pi(f_{cav}-f) - i\kappa/2 - \frac{g_{ij}^2}{-i\Gamma_{ij}/2 + \Delta_{ij}}} - i\sqrt{T}e^{i\zeta}$$

with $\Delta_{ij} = 2\pi(f_{ij} - f)$, $f_{ij}$ being the double quantum dot transition frequency, $f_{cav}$ being the resonance frequency of the cavity and $f$ being the frequency of the cavity drive. The parameter $\alpha = \sqrt{\kappa_1\kappa_2}$ accounts for the coupling capacitance of the resonator and $T$ and $\zeta$ control the Fano line shape of our resonance, which is slightly visible in figure 3B.

The above formula is the one used to fit the transmission of the cavity as shown in figure 3B in the main text. First, the cavity parameters ($\alpha, \kappa, f_{cav}, T, \zeta$) are extracted when the double dot is strongly detuned (all the $\chi_{ij}=0$). After determining the resonance point $\Delta_{ij} = 0$, we fit the data with two free parameters, $g_{ij}$ and $\Gamma_{ij}$. This allows us to extract both the cooperativity and the dephasing rate of the transition $ij$. In addition, whereas the Fano line shape is important to obtain a quantitative fit of the resonances both tuned and detuned, letting $T$ to zero does not give markedly different cooperativies and decoherence rates (in this case one has to correct for the background of the data). The amplitude $A$ is defined as $|S_{21}|$ and the phase $\varphi$ as $\arg(S_{21})$.

It is important to stress that the cooperativity controls the maximum of the resonance whereas the decoherence rate controls the width. Hence, the cooperativity can also be directly extracted from the ratio of the transmission at resonance in the tuned and detuned condition. Indeed, one can show that the above formula leads to:

$$C_{ij} = \frac{4g_{ij}^2}{\kappa \Gamma_{ij}} \approx \frac{A_{OFF}}{A_{ON}} - 1$$

Applying this formula yields C=2.3, in very good agreement with the full fitting procedure for the data in figure 3B. The approximate equality stems from the slight Fano line shape and becomes exact if $T=0$.

Theory of our experimental findings

We further support our experimental findings by a microscopic modeling of our nanotube based spin/photon coupling scheme. The starting point is the full microscopic Hamiltonian of the carbon nanotube based double quantum dot with non-collinear ferromagnetic contacts, projected onto the (1,0)-(0,1) charge states, and coupled to a single mode of the cavity. Since the double quantum dot which we study is made out of a single wall carbon nanotube, we must include the K/K' valley degree of freedom in the description (not represented in figure 2D for the sake of simplicity). The full Hamiltonian reads :

$$H_d =$$
$$-\frac{(1+\hat{\tau}_Z)}{2} \frac{(\vec{\delta}_{L,K}(1+\hat{\gamma}_Z) + \vec{\delta}_{L,K'}(1-\hat{\gamma}_Z) + \alpha_{spin}^L \mu_B \vec{B}_{ext})}{2} \cdot \vec{\hat{\sigma}}$$
$$-\frac{(1-\hat{\tau}_Z)}{2} \frac{(\vec{\delta}_{R,K}(1+\hat{\gamma}_Z) + \vec{\delta}_{R,K'}(1-\hat{\gamma}_Z) + \alpha_{spin}^R \mu_B \vec{B}_{ext})}{2} \cdot \vec{\hat{\sigma}}$$
$$+ (t\hat{\tau}_x + \varepsilon\hat{\tau}_Z)\hat{\gamma}_0 \hat{\sigma}_0 - \frac{1}{2}\alpha_{orb}\mu_B B_{ext} \hat{\gamma}_z \hat{\tau}_0 \hat{\sigma}_0 + \Delta_{KK'}\hat{\gamma}_x \hat{\tau}_0 \hat{\sigma}_0$$
$$+ g_d (a+a^+)\hat{\tau}_z \hat{\gamma}_0 \hat{\sigma}_0 + \hbar\omega_{cav} a^+ a$$

with

$$\vec{\delta}_{L(R),K/K'} = \delta_{L(R),K/K'}(\cos\theta_{L(R)}\vec{u}_z + \sin\theta_{L(R)}\vec{u}_x)$$

Here, $\vec{\hat{\sigma}}$ is the spin operator, $\hat{\gamma}_i, i \in \{x,y,z\}$ are the Pauli matrices acting in the valley space and $\hat{\tau}_i, i \in \{x,y,z\}$ are the Pauli matrices acting in the L/R space of the double quantum dot. The unit vector in the direction $x(z)$ is denoted by $\vec{u}_{x(z)}$. The vectors $\vec{\delta}_{L,K}, \vec{\delta}_{L,K'}, \vec{\delta}_{R,K}$, $\vec{\delta}_{R,K'}$ are the valley/dot dependent effective magnetic fields induced by the ferromagnetic contacts. We define the detuning $\varepsilon$ and the hopping constant $t$ between the left(L) and the right(R) dots. We use effective spin and orbital Landé factors $\alpha_{spin}^{L(R)}$ and $\alpha_{orb}^{L(R)}$. We assume that there can be a small disorder induced valley mixing which we include in the usual way with a phenomenological parameter $\Delta_{KK'}$. Finally, the interaction between the cavity photons and the double quantum dot is characterized by the coupling strength $g_d$. The cavity frequency is $\omega_{cav}$.

Since the measurements of Figures 2 and 3 are realized with a slowly varying magnetic field, they do not show the hysteretic behavior of Figure 4. Instead, the magnetizations are relaxed for each measurement point in equilibrium positions described by angles $\theta_L$ and $\theta_R$ which vary with $B_{ext}$. For simplicity we use:

$$\theta_{L(R)} = \theta^0_{L(R)} \exp(-B_{ext}/B_0) \text{ for } B_{ext} > 0$$

$$\theta_{L(R)} = \pi + \theta^0_{L(R)} \exp(B_{ext}/B_0) \text{ for } B_{ext} < 0$$

These equations take into account that the magnetizations and the external magnetic field tend to the same direction for high values of $B_{ext}$. They also take into account that for a vanishing $B_{ext}$, the orientations of the magnetizations stick to the easy axes of the electrodes, but with a $\pi$-flip from $B_{ext} > 0$ to $B_{ext} < 0$, in order to minimize the angle with the magnetic field. This last feature is essential to reproduce the cusps occurring in the resonances of Fig. 2A and 2B for $B_{ext} = 0$.

The above Hamiltonian is a generalization of the Hamiltonian (1) of *(15)* in which the valley degree of freedom was omitted. The inhomogeneity in the direction of the effective fields ($\theta_L \neq \theta_R$) induces a mixing of the spin states and the L/R orbital states. This is the main ingredient for our artificial spin orbit interaction. Since the photons induce electron hopping between the left and the right dot, they can also induce spin flips.

We determine from the Hamiltonian the transition energies $E_{ij}=h\, f_{ij}$, and the couplings $g_{ij}$. We also include in the model decoherence rates $\Gamma_{ij}$, which include relaxation and dephasing. For simplicity we use a small transition independent relaxation rate $\Gamma_1$. In our model, decoherence is dominated by charge noise treated semiclassically up to second order in order to describe properly sweet spots. Using the above equations and the transmission formula given in the previous section, we obtain the full colorscale plots of the amplitude and phase of the microwave signal as a function of the gate detuning $\varepsilon$ and the external magnetic field $B_{ext}$ (see Figure S5). This allows us to account very well for the measurements presented in Figures 2A and 2B as well as the spectroscopic lines of Figure 3. We use the following parameters: $\omega_{cav} = 6735\, MHz$, $t = 2380\, MHz$, $\delta_{L,K} = 3135\, MHz$, $\delta_{L,K'} = 3095\, MHz$, $\delta_{R,K} = 3145\, MHz$, $\delta_{R,K'} = 3100\, MHz$, $\alpha^K_{spin}\mu_B = 2700\, MHz/T$, $\alpha^{K'}_{spin}\mu_B = 1300\, MHz/T$, $\alpha_{orb}\mu_B = 300\, MHz/T$, $\Delta_{KK'} = 28\, MHz$, $g_d = 45\, MHz$, $\Gamma_1 = 1\, MHz$, $B_0 = 1,5\, T$, $\theta^0_L = -0.17\, rad$, and $\theta^0_R = \theta^0_L + \frac{\pi}{4}$. We show in Figure S4 the DQD eigenenergies obtained from Hamiltonian $H_d$ and the corresponding transitions energies $E_{ij}$, versus $B_{ext}$. The number of eigenstates is twice larger than expected from figure 2D, due to the inclusion of the K/K' degree of freedom. The two lowest eigenstates 0 and 1 of $H_d$ are only slightly split due to the slight asymmetry between the K and K' valleys. With the above parameters and the $B_{ext}$ values of the experiment, this splitting is smaller than temperature. Therefore, we assume that these two states are equally populated. From Figure S4B, the transitions 04, 15 and 25 become resonant with the cavity. They reproduce well the transitions Spin, Spin' and the third faint transition of the main text, respectively.

With the above model, the doublets in figure S5C and D arise from the slight asymmetry between the K and K' valleys. Note that the coupling between the two valleys remains very small here. With our parameters, the two strong resonances from the doublet (04 and 15) are dominated by spin-flip and (to a weaker proportion) by L/R flips. These two transitions are mainly valley conserving. Their contrast is rather well reproduced by our simulation, which confirms that decoherence of our spin states is indeed to a great extent caused by charge noise, due to the spin/charge hybridization. In particular, we can reproduce the existence of a minimum of decoherence and a maximum in the spin-photon cooperativity near $\varepsilon = 0$. This behavior is due to the existence of a sweet spot with respect to charge noise, indicated by the green dotted line in Fig.S5C. In constrast, the third faint resonance (25) has a more important K/K' flip component. The contrast of this resonance is stronger in the calculation than in the data, which may be corrected by introducing a specific intrinsic K/K' relaxation rate. Indeed, the K/K' degree of freedom is expected to be intrinsically less coherent than the spin degree of freedom because it can be directly affected by spin-conserving decoherence sources such as phonons.

Note that in our experiment, we could not determine the parity of the DQD states. We have chosen to use a one electron model because it reproduces well the data. It is nevertheless important to point out that in principle, the parity of the electron states should affect only quantitatively the behavior of our device. Indeed, the concept of an artificial spin-orbit coupling induced by non-collinear ferromagnetic contacts remains valid for even occupation states. The robustness of this principle is a significant advantage of our scheme.

Hysteresis measurements

We use a standard definition of the magneto-conductance $\frac{I_{DC}^{incr} - I_{DC}^{decr}}{I_{DC}^{incr} + I_{DC}^{decr}}$. Because the phase signal can have both sign, we normalize the phase variations slightly differently to avoid large divergences when transitions are resonant with the cavity. We define the magneto-phase as $\frac{\Delta\varphi^{incr} - \Delta\varphi^{decr}}{\left|\Delta\varphi^{incr}\right| + \left|\Delta\varphi^{decr}\right|}$.

It is important to check that the hysteresis we observe in the phase of the cavity is due to the susceptibility of the coupled spin transitions and does not arise from spurious hysteretic behavior of the bare cavity mode. Figure S6 shows the bare superconducting cavity characteristics for increasing and decreasing magnetic field, and their hysteresis in percent. The presented data are extracted during the same magnetic field cycle as in figure 4 of the main text. They are acquired for far detuned double quantum dot transitions, as explained above. In the main text, our purpose is to detect the phase variation caused by the DQD, which is given by Δφ ~ (phase slope)×Re($\chi_{ij}$) at first order in the DQD/cavity coupling. In figure 4D of the main text, the observed hysteresis of the phase shift reaches more than 50% at small fields and up to 100% at DQD/cavity resonances. In contrast, the bare cavity parameters all show a small hysteresis, from less than 0.1% for the resonant frequency to few percent for cavity linewidth and phase slope at resonance, with qualitatively different variations from figure 4D. Therefore the data of the main text are not due to a simple hysteresis from the cavity.

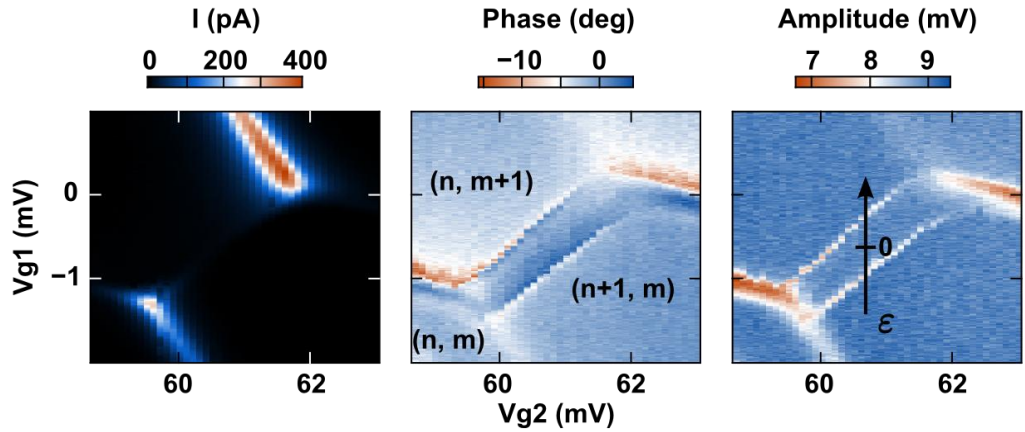

**Fig. S1.**
**Stability Diagram.** Double quantum dot transport spectroscopy showing the region on the stability diagram studied in the paper ($V_{gt}$ = -1mV, $B_{ext}$ = 0mT). Left panel shows a color-scale plot of the DC current ($V_{SD} \approx 20\mu V$). Middle and right panels show phase and amplitude of the transmission coefficient $S_{21}$, see SM text. Stable charges states are labeled by (n,m) with n and m occupation numbers in each dot.

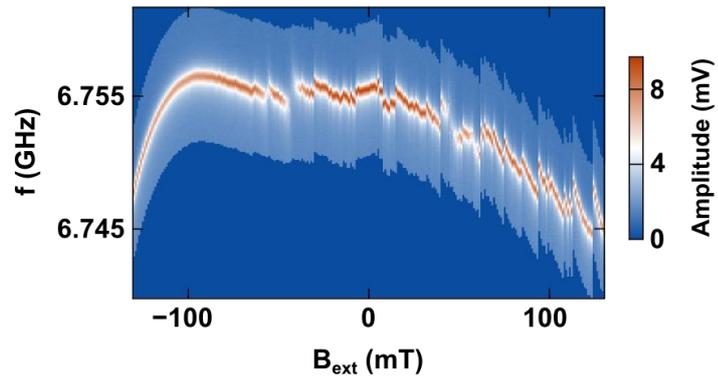

**Fig. S2**
**Cavity transmission in absolute frequency.** Measured transmission spectrum of the cavity as a function of $B_{ext}$ without post-treament.

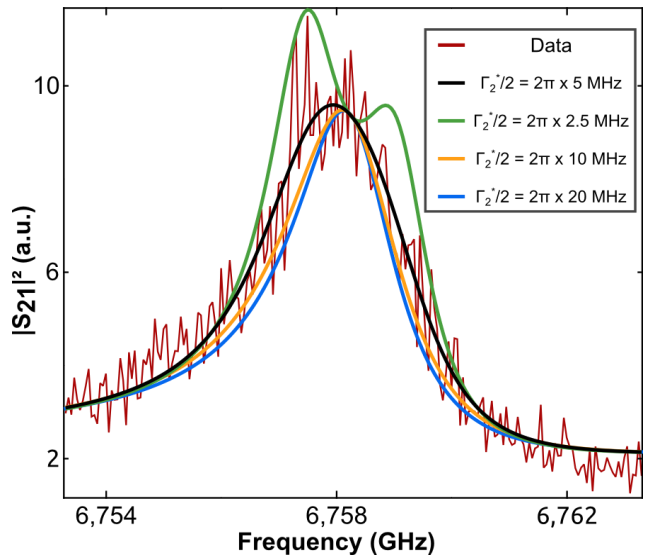

**Fig. S3**
**Extracting the decoherence rate.** Transmission of the cavity at -67 mT, when the spin transition is brought in resonance with the cavity (as described in the main text). The black curve is the best fit and the other curves correspond to different spin decoherence rates.

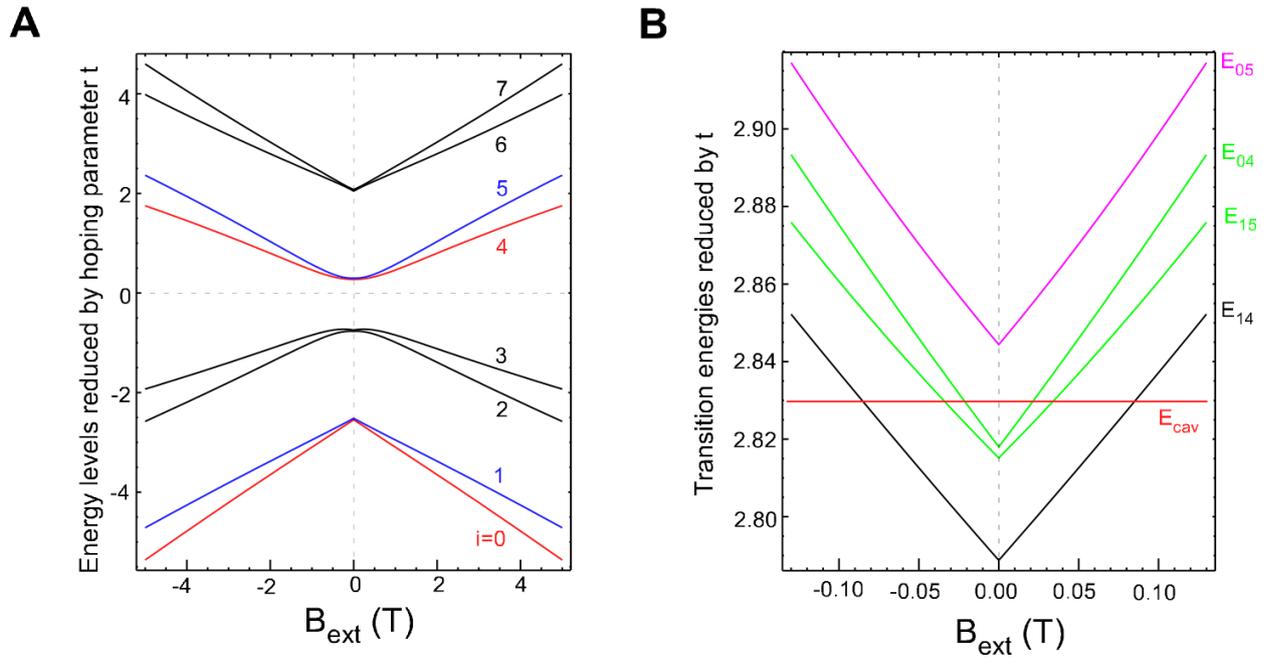

**Fig. S4**
**Spectrum obtained from the theoretical model.** (A) and (B) Calculated DQD energy spectrum and transition energies $E_{ij}$ versus $B_{ext}$, for $\varepsilon=0$ and the parameters given in the section "Theory of our experimental findings". The energy levels are labelled with an index i. For the $B_{ext}$ values used experimentally, the transitions energies $E_{04}$, $E_{15}$, and $E_{04}$ can become resonant with the cavity. We have only represented the transitions which are close to the cavity for the sake of clarity.

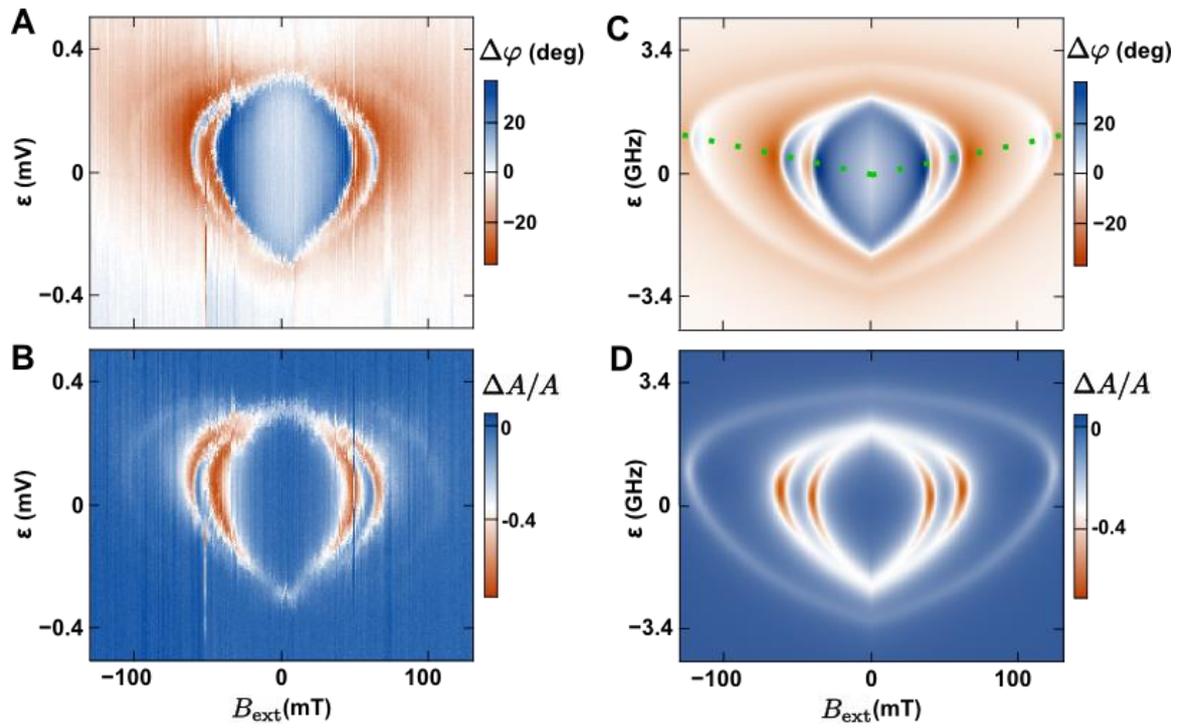

**Fig. S5**
**Electric and magnetic dependence of the transitions: experiment and theoretical model.** (A) and (B) phase and amplitude measured as in the main text. (C) and (D) modelling of the phase and amplitude using the microscopic Hamiltonian described in the SM. The green dotted line in panel C corresponds to a sweet spot with respect to charge noise.

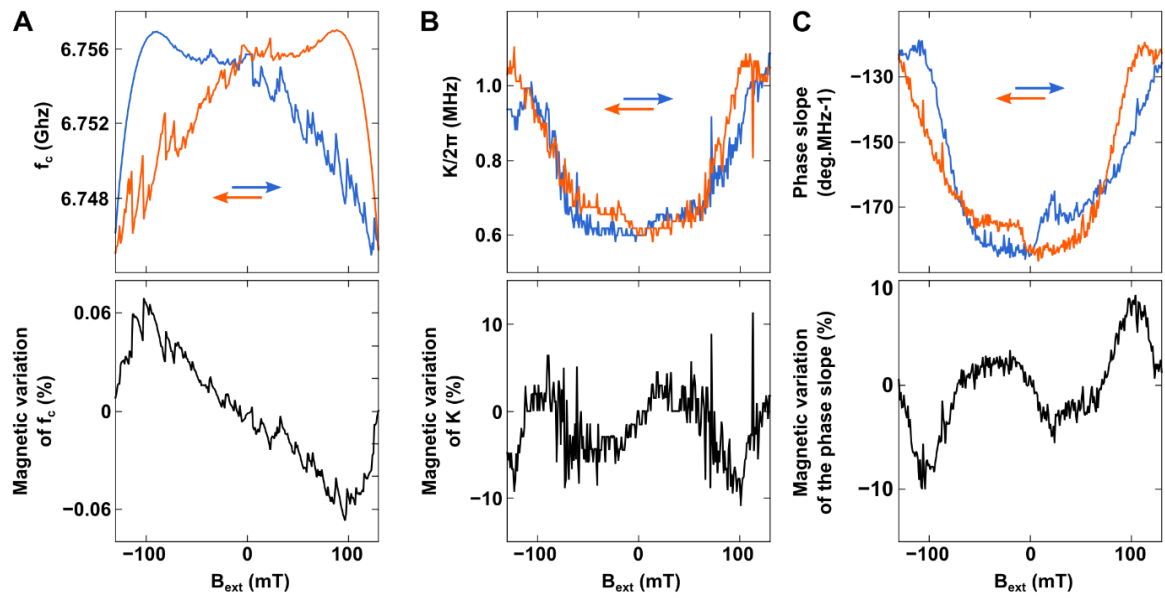

**Fig. S6**
**Hysteresis of the cavity properties.** Bare superconducting resonator characteristics (double quantum dot transitions far detuned) as a function of magnetic field and magnetic field sweep direction. (A) Resonant frequency. (B) Cavity linewidth. (C) Phase slope at resonance.